\documentclass[pdflatex,sn-mathphys-num]{sn-jnl}

\usepackage{graphicx}%
\usepackage{multirow}%
\usepackage{amsmath,amssymb,amsfonts}%
\usepackage{amsthm}%
\usepackage{mathrsfs}%
\usepackage[title]{appendix}%
\usepackage{xcolor}%
\usepackage{textcomp}%
\usepackage{manyfoot}%
\usepackage{booktabs}%
\usepackage{algorithm}%
\usepackage{algorithmicx}%
\usepackage{algpseudocode}%
\usepackage{listings}%
\usepackage{todonotes}
\usepackage{graphicx}
\usepackage{subcaption}

\theoremstyle{thmstyleone}%
\theoremstyle{thmstyletwo}%

\theoremstyle{thmstylethree}%

\begin{document}

\title[Article Title]{The geopolitics of knowledge: tipping points, national fingerprints, and the unequal globalization of science}


\author[1]{\fnm{Irina} \sur{Vorobeva}}\email{irina.b.vorobeva@gmail.com}
\equalcont{These authors contributed equally to this work.}

\author[2]{\fnm{Maxime} \sur{Lenormand}}\email{maxime.lenormand@inrae.fr}

\author[3]{\fnm{Germana} \sur{Berlantini}}\email{germana.berlantini@gmail.com}

\author*[1]{\fnm{Floriana} \sur{Gargiulo}}\email{floriana.gargiulo@cnrs.fr}
\equalcont{These authors contributed equally to this work.}

\affil*[1]{\orgdiv{Gemass}, \orgname{CNRS}, \orgaddress{\street{59 rue Pouchet}, \city{Paris}, \postcode{75018},  \country{France}}}

\affil[2]{\orgdiv{TETIS}, \orgname{University of Montpellier, AgroParisTech, Cirad, CNRS, INRAE}, \orgaddress{\city{Montpellier}, \country{France}}}

\affil[3]{\orgdiv{Centre des politiques de la Terre}, \orgname{Université Paris Cité}, \orgaddress{\city{Paris}, \country{France}}}


\abstract{Science is often portrayed as a universal and self-contained system, driven solely by the internal logic of knowledge accumulation and isolated from the turbulences of the socio-political world. In this paper, we challenge this narrative by providing systematic quantitative evidence that the global scientific ecosystem is deeply shaped by geopolitical transformations. Using a large-scale dataset of scientific publications drawn from the OpenAlex database, spanning over five decades and covering virtually all countries and disciplinary areas, we track the evolution of national research profiles and show that geopolitical dynamics shape scientific agendas at multiple scales. At the global level, intrinsic scientific change is slow and cumulative, but exogenous shocks---such as Chernobyl, September 11, and COVID-19---produce rapid disruptions that synchronously reconfigure the priorities of many countries at once. At the country level, we document a broad globalization of knowledge, yet deeply heterogeneous: while Global North countries converge toward a shared international agenda, Global South countries display strong dependence on international resources alongside locally distinctive research interests. Among emerging Southern economies, scientific power is increasingly asserted through specialized and independent agendas. Finally, we observe a reorganization of global scientific influence toward a more polycentric structure, with the emergence of a Southern cluster gravitating around Brazil and Indonesia as new regional hubs.

}



\maketitle

\section{Introduction}\label{sec1}
Science is commonly described as a cumulative process of knowledge production, in which each generation of researchers builds upon the work of previous ones to gradually expand the scientific ecosystem: the famous concept of ``standing on the shoulders of giants''. This image of science as a self-reinforcing, internally driven system has deep roots in the philosophy and sociology of knowledge, from Merton's norms of universalism and communalism \cite{merton1973sociology}  to more recent scientometric literature on cumulative knowledge production \cite{della2020tracking}. The opposing view is that of Kuhn \cite{khun1970structure}, who argued that science does not progress smoothly but rather through abrupt paradigm shifts: periods of crisis in which an established theoretical framework is overthrown and replaced by an entirely new one, introducing fundamental discontinuities in the development of scientific knowledge. Several papers debated on these opposing visions \cite{krauss2024debunking,newig2026rethinking,della2020tracking}, also using scientometric indicators like the ``disruptivness'' \cite{park2023papers}.

The hypothesis of  cumulative science holds well at the microscopic level. At the scale of individual researchers, career trajectories are largely cumulative: scientists tend to deepen and extend their expertise over time, combining their own prior work with that of their intellectual community \cite{sinatra2016quantifying,foster2015tradition}. The description of science discovery as a recombination process is present in several studies \cite{uzzi2013atypical} and is at the basics of the Kaufmann's description of innovation processes as progressive widening of the adjacent possible \cite{kauffman1995home}.
However, aggregation does not always preserve regularity. This is a well-known phenomenon in complex systems theory: macro-level dynamics can be qualitatively dissimilar from the micro-level trajectories that generate them, as collective behavior gives rise to emergent properties that are not present in and are not predictable from the properties of individual components. When scientific production is measured at the national level, macroscopic dynamics can exhibit emergent phenomena that are not predictable from individual behavior alone. National research agendas are shaped not only by the internal logic of knowledge accumulation, but also by institutional priorities, funding structures, geopolitical forces and local cultures.

The idea that science is deeply entangled with power and politics is far from new. It has been a central theme of numerous ideological and intellectual movements throughout the twentieth century, from Marxist critiques of science \cite{ciccotti1976ape,hessen2009social,needham1954china}, to Foucauldian reflections on the relationships between power and knowledge \cite{foucault2002archeology}, and to Kuhnian accounts of how paradigms reflect the social context of their time \cite{khun1970structure}. Philosophers and sociologists of science have long argued that national scientific agendas are not neutral entities, disconnected from economic, political or geopolitical relations. Rather, the questions researchers choose to pursue,  whether at the level of national scientific systems or transnational research communities, and the methods they regard as legitimate are all embedded in broader structures of power, that influence what counts as relevant knowledge, which problems deserve investigation, and whose perspectives are recognized as authoritative \cite{keller1985gender,jasanoff2004states,harding2008below}.

This argument has been developed with particular force, in recent decades, by decolonial scholars, who have drawn attention to the asymmetric power relations between the Global North and the Global South in the production of scientific knowledge \cite{alatas2003dependency,hountondji1990africa,mignolo2002geopolitics}. Contrary to the image of a universalist and value-free endeavor, global science tends to marginalize Southern scientific production or to subordinate it to research agendas defined in the North, thereby neglecting local epistemic needs and priorities \cite{connell2020southern,collyer2019power}. North-South scientific collaborations seem to reinforce this hierarchical structure, often assigning Southern partners the task of empirically validating theories and hypotheses formulated in the North or providing fieldwork expertise, thereby reproducing a global division of academic labor in which Southern intellectual resources are mobilized in support of research agendas, priorities, and purposes largely defined elsewhere \cite{kreimer2008connaissance, collyer2019power}. The same exploitation pattern has also been observed for the Arctic Region \cite{strouk2024field}.

Another concept that has emerged in the last decades is, indeed, that of ``globalization of science'' \cite{gui2019globalization}.
Since the late twentieth century, the acceleration of economic and cultural globalization has left a clear imprint on scientific production, most visibly in the dramatic rise of international collaborations \cite{ribeiro2018growth, wagner2017growth}. Yet most of this evidence concerns the \textit{internationalization} of science, namely the growing tendency of researchers to co-author and being mobile across borders. 
Several authors have shown that international mobility \cite{gargiulo2014driving, sinatra2016quantifying} and co-authorship \cite{ductor2015does} are predictors of scientific impact (in terms of citation accumulation). At the same time, it emerges that such dynamics are at the origin of the amplification of intrinsic inequalities in science production, above all in relationship to gender \cite{macedo2023academic,huang2020historical}. 

Developing in the computational framework of science of science \cite{fortunato2018science}, in this paper, we move beyond collaboration and citation networks to study directly the scientific agenda of nations: what countries choose to study, and how these choices evolve over time and in response to external pressures.

A substantial body of bibliometric work has characterized national scientific profiles and compared them across countries, revealing clustering patterns that reflect long-term policies, degree of development, and geographical proximity as proxies for epistemic alignment \citep{ScienceIndicators2009, DiversitySimilarity2012}.

Few studies ask the deeper question of whether national scientific agendas are converging or diverging \cite{cimini2014scientific}. Collaboration can increase without homogenization: a country may co-author extensively with foreign partners while pursuing a domestically distinctive research agenda, or conversely, it may align its priorities with global trends without ever co-authoring internationally.
Longitudinal analyses have produced a nuanced picture: \citet{EvolutionaryPatterns2017} find a continuous convergence trend across G7 and BRIC nations, while \citet{Convergence2011} show that on a broader set of countries convergence proceeds within clubs rather than globally, and propose socio-historical explanations for the observed groupings.

Across all of these studies, however, domestically produced papers and internationally co-authored ones are pooled into a single national output vector, conflating a country's endogenous research priorities with the agenda of its foreign partners. Separating the two opens broader questions about dependency and the structural position of national research systems in the global landscape, a comparison that, to our knowledge, has not yet been undertaken.

In this work, we construct epistemic profiles for each country: vectors describing the distribution of domestic scientific production across disciplines. We track their evolution over five decades using a large-scale dataset of publications drawn from the OpenAlex database \cite{openalex}, covering virtually all countries and disciplinary areas.

Our analysis yields five main findings. First, epistemic profiles are sensitive to major geopolitical shocks: events such as the Chernobyl disaster, the September 11 attacks, and the COVID-19 pandemic produce rapid and synchronized reconfigurations of scientific priorities across many countries, standing out against the otherwise slow and cumulative pace of scientific change.

Second, we document a broad trend toward the internationalization of scientific production, but reveal a structural asymmetry: while Global North countries show increasing alignment between their domestic and international research agendas, Global South countries exhibit a systematic divergence between the two. In the South, international production, often driven by collaborations, tends not to reproduce the domestic scientific agenda, suggesting that Southern intellectual resources are frequently mobilized in service of externally defined priorities, a process that has been conceptualized as “extraversion” \cite{hountondji1990africa}.

Third, we find that while dominant scientific powers tend to spread their production broadly across all disciplines, emerging countries assert their scientific identity through specialization, carving out distinctive and focused research agendas rather than competing across the board.

Fourth, the structure of scientific aspiration (how countries orient themselves toward foreign scientific models) is undergoing a significant reorganization. The global landscape of scientific influence is becoming more polycentric, with new regional hubs emerging alongside the traditional dominance of North American and European science.

Fifth and finally, we identify the consolidation of a Southern scientific cluster, gravitating around Brazil and Indonesia, which is distinguished by a shared focus on social sciences and agriculture. This cluster represents not just a grouping of similar countries, but a coherent and increasingly autonomous scientific community with its own thematic priorities.

Together, these findings paint a picture of a  scientific world-system that is far from the self-evolving science idea: deeply stratified, responsive to political shocks, and structured by the same geopolitical asymmetries that characterize the broader world order.


\section{Results}
\subsection{Tipping points}

\begin{figure}[h]
\centering
\includegraphics[width=0.95\linewidth]{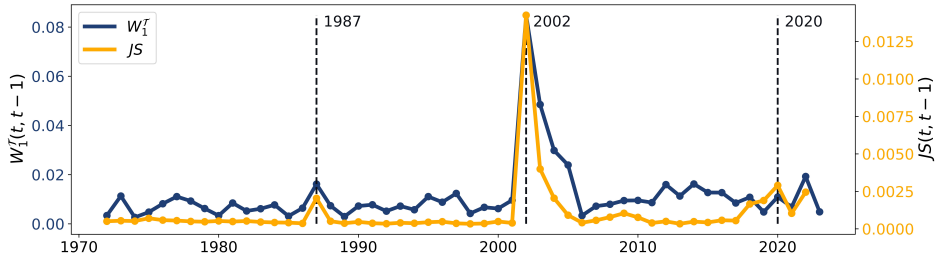}
\caption{Profile distances between two subsequent years, measured with JS divergence and $W_1^{\mathcal{T}}$ distance. We can identify 3 picks: in 1987, 2002, and 2020.}\label{fig1}
\end{figure}

We first analyze the temporal evolution of the worldwide-aggregated domestic research profile, the share of each discipline in the annual scientific production, described by Eq.\ref{worldprofiles}. We use two different measures to evaluate the year-to-year variations: the Jensen-Shannon divergence and the tree-Wasserstein distance. The first captures the intensity of redistribution, and the second captures the depth. 
As we can observe in Fig.\ref{fig1}, the two metrics agree on a main question: the evolution of research profiles is characteristically slow under normal conditions. In the absence of external perturbations, the global distribution of scientific output across disciplines changes gradually, reflecting the incremental nature of knowledge accumulation. Against this backdrop of endogenous, slow-moving change, the two metrics reveal three sharp and anomalous disruptions, coinciding with major exogenous shocks.

\begin{figure}
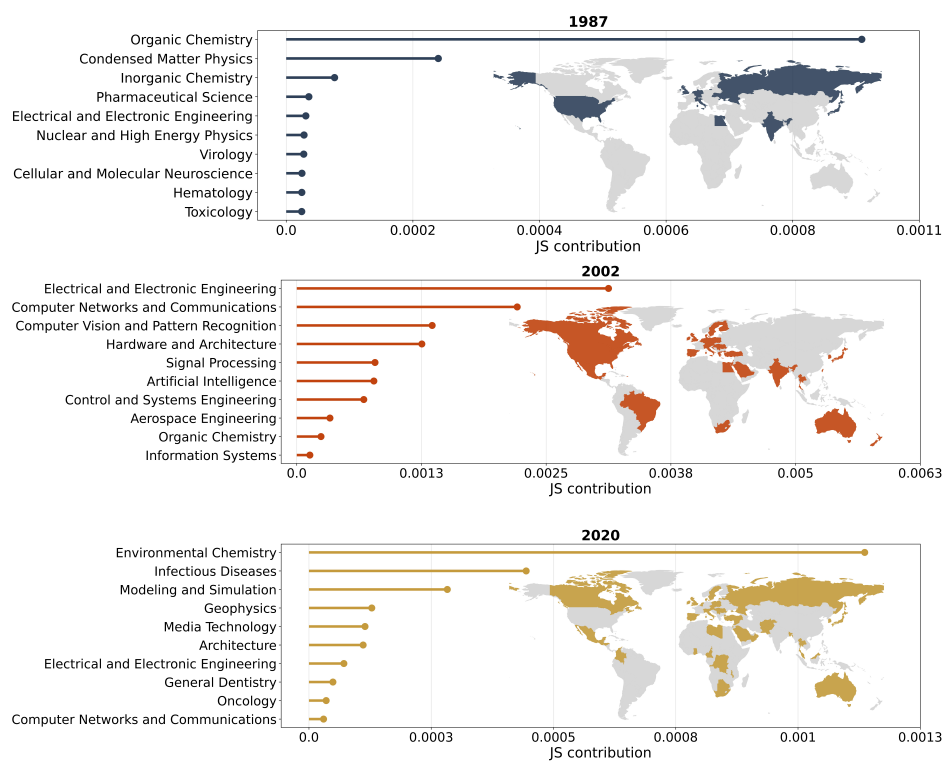

\centering
\includegraphics[width=0.95\linewidth]{figures/map_js_contribution_1987_v1.jpg}
\vspace{0.3cm}
\includegraphics[width=0.95\linewidth]{figures/map_js_contribution_2002_v1.jpg}
\vspace{0.3cm}
\includegraphics[width=0.95\linewidth]{figures/map_js_contribution_2020_v1.jpg}
\caption{Characteristics of the three main picks: disciplines driving the shift and countries involved in the change. The bright zones indicate the occurrence of notable shifts in topic prioritization.}
\label{fig1b}
\end{figure}

The discipline-level decomposition of the JS  divergence, Fig.\ref{fig1b}, provides a fingerprint of each exogenous shock. By identifying which disciplines drive the aggregate peak in each disruption year, we can trace the pathway through which external events translated into a reorganization of scientific priorities, moving beyond the observation that a shock occurred to understand where in the disciplinary space it left its mark.

The peak around 1987–1988 aligns with the Chernobyl disaster (1986). The discipline-level decomposition reveals a coherent picture: the transformation is driven not only by Nuclear and High Energy Physics but also by a cluster of biomedical and chemical disciplines (Toxicology, Virology, Hematology, and Pharmaceutical Science) reflecting the broad scientific mobilization triggered by the health and environmental consequences of the accident. This pattern reveals a massive research effort directed at understanding radiation exposure, contamination, and its effects on human health and ecosystems in the years immediately following the disaster.

The most dramatic disruption occurs in 2001–2003, where both metrics reach their global maximum. The discipline-level decomposition provides a remarkably coherent fingerprint of the post-9/11 security shock: the transformation is driven by a set of defense and surveillance-related fields, including Electrical and Electronic Engineering, Computer Networks and Communications, Signal Processing, Hardware and Architecture, Control and Systems Engineering, and Aerospace Engineering, all core components of military and homeland security technology. The presence of Artificial Intelligence and Computer Vision and Pattern Recognition further reflects the rapid expansion of surveillance, threat detection, and autonomous systems research that followed the attacks. Notably, a finer-grained analysis of the AI discipline reveals that the growth was concentrated specifically in expert systems, a paradigm that had been in steady decline through the 1990s following the so-called "AI winter", and which experienced a localized revival driven by their potential applications in military objectives \cite{gargiulo2023meso, cardon2018revanche}. This pattern is consistent with the start of a decade-long increase in defense-related research funding among OECD countries \cite{oecd2026rd} that reshaped scientific priorities across the physics and engineering community in the years immediately following September 11, temporarily reversing trends that had been consolidating for over a decade. 
The divergent post-peak behavior of the two metrics adds a further layer of interpretation: while the JS divergence returns rapidly to baseline levels after 2003, suggesting that the identity of dominant disciplines stabilized relatively quickly, the $W_1^{\mathcal{T}}$ distance decays much more slowly, revealing that the reorganization involved disciplines that are structurally distant in the disciplinary hierarchy. The defense-driven fields that surged after 9/11 were not adjacent to the previously dominant areas of physics research, and their gradual reabsorption into the broader landscape required years of structural realignment, a process that is invisible to JS but captured by the tree-sensitive geometry of $W_1^{\mathcal{T}}$.

Finally, elevated and volatile values after 2020 reflect the impact of the COVID-19 pandemic. The discipline-level decomposition reveals a dual signature: on one hand, the expected surge in directly pandemic-related research, with Infectious Diseases and Modeling and Simulation capturing the massive mobilization around epidemiological modeling and outbreak forecasting; on the other, a broader reorganization driven by the structural disruptions that the pandemic imposed on society at large. The emergence of Environmental Chemistry, for example, likely reflects the temporary but measurable impact of lockdowns on pollution and environmental monitoring, while Computer Networks and Communications and Media Technology mirror the abrupt shift toward remote work, digital infrastructure, and online communication. More surprisingly, the presence of Architecture among the top contributing disciplines points to the profound restructuring of physical spaces triggered by the pandemic.

We also analyzed the geographic patterns of these transformations, identifying which countries experienced an anomalous peak in the shock years.

In 1987, the set of affected countries is small and geographically concentrated: Russia and Ukraine representing the USSR, the most directly exposed to the Chernobyl disaster, show anomalous peaks, alongside a handful of major Western European scientific powers (Germany, Italy, Switzerland, the Netherlands, the United Kingdom) and Japan. The limited geographic footprint of this shock reflects its regional nature: while Chernobyl had global resonance, the scientific response was concentrated in countries with direct exposure to the fallout or with strong nuclear research traditions.

The 2002 shock presents a dramatically different picture. The list of affected countries expands to over forty nations, spanning Europe, North America, Asia, the Middle East, and Latin America. This near-global reach is consistent with the interpretation of a funding-driven shock: the post-9/11 restructuring of research priorities propagated through international funding networks and defense partnerships, touching virtually every country with a substantial scientific community. Notably absent are China and Russia, suggesting that the reallocation of defense-related research funding was primarily a Western and US-aligned phenomenon.

The 2020 shock is the most geographically diffuse of the three, encompassing not only the most advanced scientific nations, but also a large number of lower-income countries, including several African nations (Ghana, Cameroon, Democratic Republic of Congo, Malawi, Burundi, Botswana), Central Asian countries (Afghanistan, Pakistan, Bangladesh), and smaller economies across Southeast Asia and Latin America. This broadening of the geographic footprint reflects the truly global and indiscriminate nature of the COVID-19 disruption, which perturbed research activity regardless of a country's scientific capacity or geopolitical alignment. Interestingly, the United States did not appear among affected countries in 2020. 

\begin{figure}[h]
\centering
\includegraphics[width=0.95\linewidth]{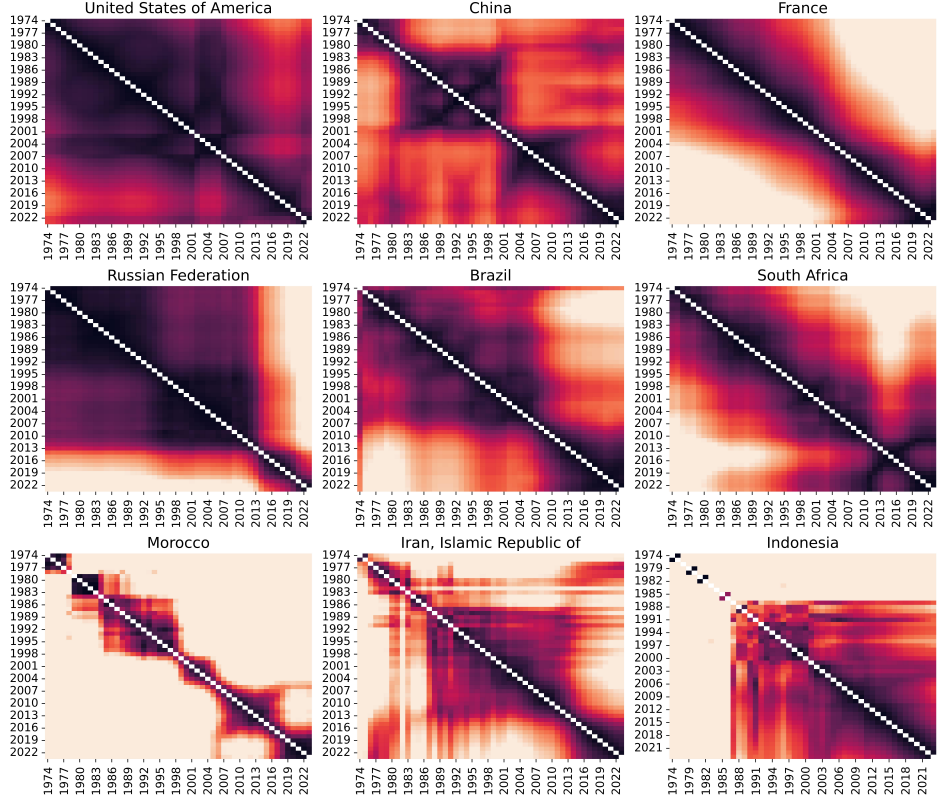}
\caption{Stability matrix: for each country we present the distance heatmap of scientific profiles across the years of analysis. The darker is the color, the closer the profiles are.}\label{fig1c}
\end{figure}

After observing the world scale agenda,  we take a closer look at how these dynamics manifest at the country level by examining shifts in domestic scientific agendas. For each country, we calculate the tree-Wasserstein distance between each pair of years in our analysis (Fig.\ref{fig1c}). This allows us not only to compare consecutive years, but also to trace the ripple effects of external shocks on science over the long term.

We first observe that the overall stability of scientific agendas varies considerably across countries. The US scientific agenda shows well a large stability except for the major change around the year 2001, due to the effect of 9/11. Russia's heatmap shows sudden but relatively contained changes following the dissolution of the Soviet Union in 1991, and an unprecedentedly rapid shift in 2014. The later 2021 shift drove the domestic profile even further from Soviet agenda. France's domestic scientific profile shows a gradual but substantial transformation.
One of the most interesting cases is China, whose scientific profile shifts periodically and substantially, reflecting the role of the Communist Party in shaping the national research agenda. The five-year blocks visible in the 1980s and 1990s suggest the influence of political five-year planning cycles on Chinese science. Brasil stability matrix shows a significant shift in 2007, coinciding with the major political interventions of the President Luiz Inácio Lula da Silva for scientific research (PACTI – Plano de Ação em Ciência, Tecnologia e Inovação). 

Taken together, these results challenge the conventional narrative of science as a self-contained, politically neutral enterprise driven solely by internal intellectual dynamics. The structure of global research, which disciplines are pursued, by whom, and where, is deeply shaped by the geopolitical landscape. Major external shocks do not merely perturb the scientific community at the margins; they fundamentally reorganize its priorities, redirect its resources, and redraw its geography, often in ways that persist for years after the triggering event. Far from being insulated from the world, science is embedded in it.

\subsection{Unequal globalization}
Among the most consequential geopolitical transformations of the late twentieth century is the rise of globalization. Broadly understood as the intensification of cross-border flows of goods, capital, people, and information, it accelerated dramatically following the end of the Cold War, reshaping the structure of international relations, economic interdependence, and institutional cooperation. To what extent did these forces also penetrate the organization of scientific knowledge production? 

To address these questions, we analyze the globalization of science through the lens of collaborative output. We first examine whether international co-authorship has grown over time --- that is, whether an increasing share of scientific production results from collaborations that cross national borders. 

\begin{figure}[h]
\centering
\includegraphics[width=0.95\linewidth]{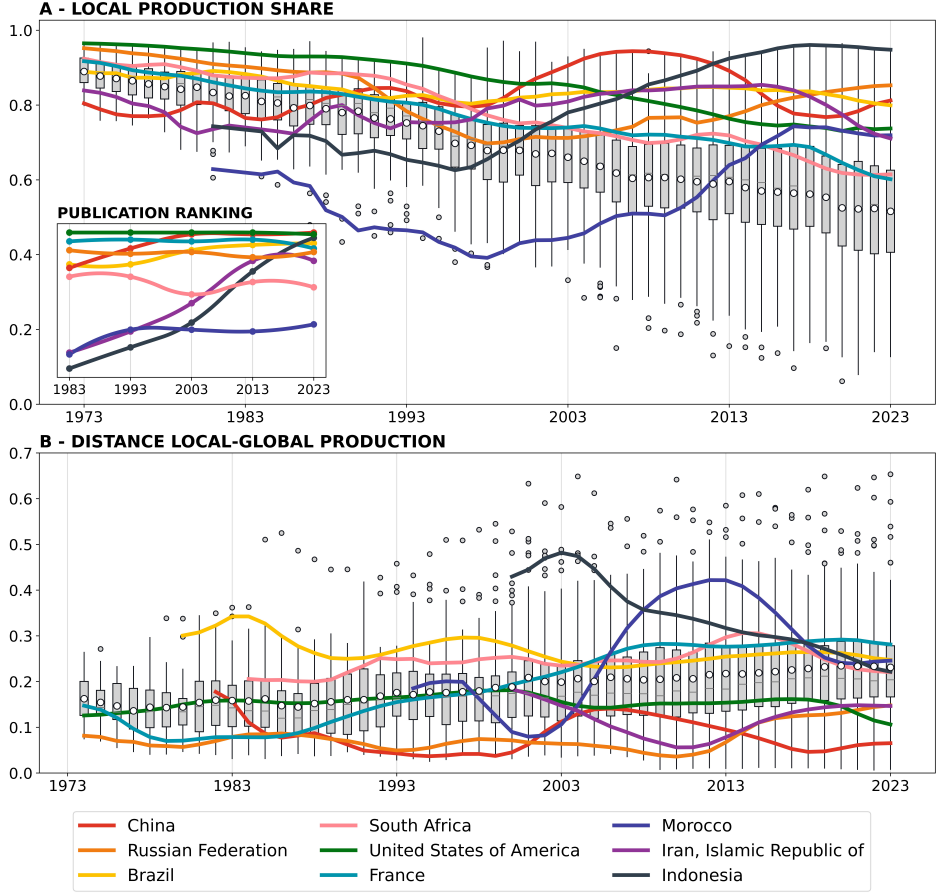}
\caption{(A) Evolution of the domestic share of scientific output from 1973 to 2023. The boxplot distribution summarizes the cross-country variation at each year; colored lines track individual country trajectories for a selected set of nations. The inset shows the corresponding evolution of publication ranking. (B) Evolution of the $W_1^{\mathcal{T}}$ distance between each country's domestic research profile and its international co-authorship profile.} \label{fig_domestic_global}
\end{figure}

As we can observe in Fig.\ref{fig_domestic_global}, a the purely domestic output, namely papers produced without any international co-authorship, has declined steadily across the board since the 1970s, consistent with a broad trend toward internationalization of scientific collaboration. However, this decline is far from uniform. While countries historically at the top of the productivity ranking (Fig.~\ref{fig_domestic_global}A-inset), such as the United States and France, follow a similar declining trend, developing countries such as Morocco, Iran, and above all Indonesia show a strengthening of domestic production, bucking the broader globalization trend. One possible interpretation is that these countries are undergoing a process of scientific emancipation: having initially built their research capacity through international partnerships, often shaped by historical ties of dependency with former colonial powers or dominant scientific nations, they may now be consolidating a more autonomous national research agenda, reflecting domestic priorities and local knowledge traditions rather than the thematic preferences of external collaborators. Under this reading, the rise in domestic share is an attempt to construct an independent fingerprint that is no longer simply a peripheral reflection of the global mainstream.

Fig.\ref{fig_domestic_global}B shifts the focus from the quantity of international collaboration to its content, measuring the $W_1^{\mathcal{T}}$ distance between a country's domestic research profile and its international co-authorship profile. A striking divergence emerges over time: for most countries, what is produced internationally has become increasingly dissimilar from what is produced domestically, suggesting that international collaboration is not simply a scaled-up version of national research but reflects a qualitatively different set of priorities. Again, country-level heterogeneity is pronounced: China shows a decrease in this distance from the 2000s onward, reflecting the growing alignment of interests between domestic and international research, while France displays the opposite trend, with its international profile diverging from its domestic one over time.

\begin{figure}[h]
\centering
\includegraphics[width=0.95\linewidth]{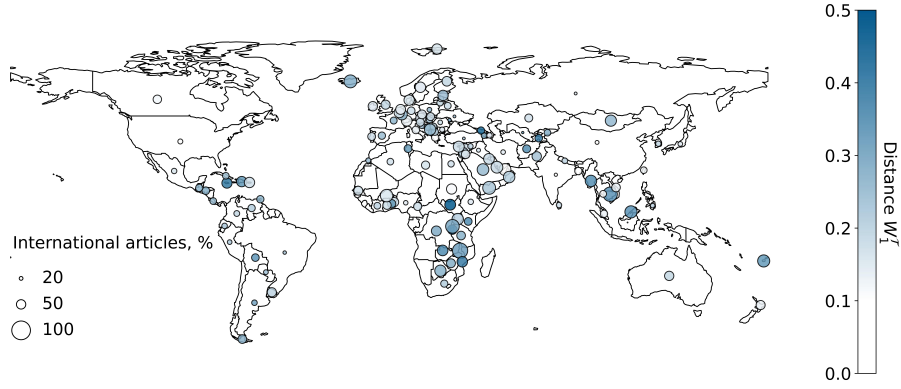}
\caption{Geographic summary of the relation between domestic and international production for 2023: circle size represents the fraction of production in international collaboration; color intensity represents the $W_1^{\mathcal{T}}$ distance between domestic and international profiles.}
\label{fig2b}
\end{figure}

Fig.\ref{fig2b} provides a geographic summary for 2023, simultaneously mapping the fraction of international production (circle size) and the distance between domestic and international profiles (color intensity). A clear global divide emerges: countries of the Global South tend to have both a lower share of domestic production and a higher divergence between their national and international profiles. This pattern suggests that for peripheral scientific communities still exists a form of scientific dependency in which international collaborations are dominant, and the content of international output is shaped more by the priorities of partners than by local needs or traditions.

\subsection{Specialization}
To characterize the breadth of each country's research portfolio, we measure its degree of specialization using two complementary entropy measures. The first is the normalized Shannon entropy, $H$, which quantifies how evenly a country's output is distributed across disciplines, treating all disciplines as equally distant from one another. The second is the Wasserstein entropy, $\mathcal{H}i = W_1^{\mathcal{T}}(p,u)$, defined as  the distance between a country's research profile $p$
and the uniform distribution $u$ over all disciplines. A high Wasserstein entropy indicates that the country's profile is far from uniform, i.e. strongly specialized in a structurally compact region of the disciplinary space; a low value indicates a profile close to uniform, i.e. broadly diversified. Used together, the two measures allow us to distinguish between qualitatively different forms of specialization: a country concentrating on a coherent disciplinary cluster will show declining entropy on both measures, while a country selecting a narrow but structurally dispersed set of fields will show declining Shannon entropy but stable or increasing Wasserstein entropy.

\begin{figure}[h]
\centering
\includegraphics[width=1\linewidth]{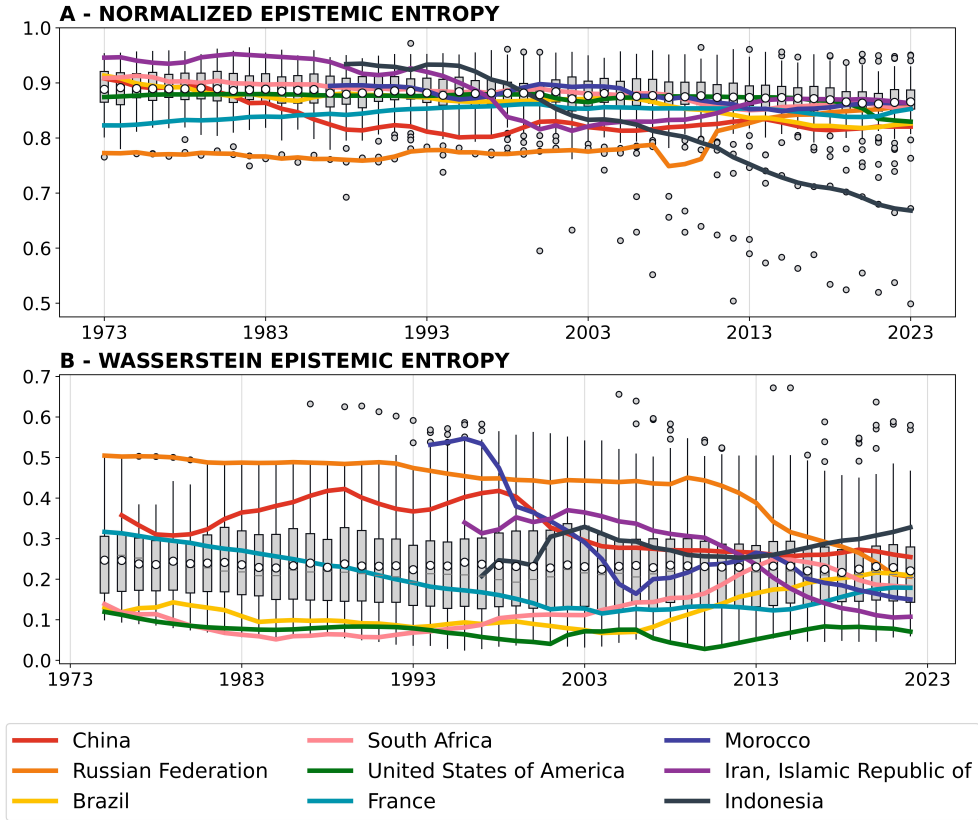}
\caption{Specialization of national research portfolios. (A) Normalized Shannon entropy of each country's research profile. (B) Wasserstein entropy, distance between a country's research profile and the uniform distribution over all disciplines.}\label{fig_entropy}
\end{figure}

Fig. \ref{fig_entropy} shows the evolution of both entropy measures from 1973 to 2023. At the aggregate level, the global distribution of entropy across countries remains remarkably stable over time on both measures, suggesting that the overall degree of specialization in the world scientific system has not changed dramatically. However, this aggregate stability conceals substantial and systematic heterogeneity at the country level, with trajectories that carry geopolitical meaning beyond internal scientific dynamics.

The United States stands out as the only country maintaining consistently low and stable entropy on both measures throughout the entire period. It is a signature of a scientific superpower whose research portfolio is broad, diversified, and evenly spread across virtually all areas of knowledge. 
Notice, however, that this measure carries a potential interpretation bias: since the US has accounted for the largest share of global scientific output over the past 50 years (23\% over the whole period), it is reasonable to expect that the underlying disciplinary classification is itself skewed toward the structure and priorities of US science.

Russia, China, and Iran follow a generalizing trajectory, with both entropy measures declining over time toward the uniform distribution. This reflects a deliberate strategic ambition to build comprehensive, self-sufficient scientific capacities spanning the full breadth of the knowledge landscape. 

France, Brazil, and Indonesia show instead a specializing trend, though with important nuances. France exhibits a moderate and gradual increase in Wasserstein entropy, consistent with a slow strategic concentration toward structurally coherent areas of established excellence. The growing divergence between domestic and international output suggests that the internal areas of focus differ from those driving France's contribution to global science. Brazil and Indonesia present the most pronounced specializing trajectory: Shannon entropy declines sharply and Wasserstein entropy rises, indicating that the countries are concentrating on a progressively smaller and structurally compact cluster of disciplines. This expresses a form of deep specialization that may reflect pragmatic choices shaped by local needs, funding opportunities, or the consolidation of a distinctive national research identity.

Finally, South Africa and Morocco exhibit a non-monotonic pattern on both measures  (an initial phase of specialization followed by a broadening of the portfolio) possibly reflecting a two-stage process of scientific development, in which limited resources initially force concentration on a narrow set of fields, before growing institutional capacity and international integration enable access to a wider range of disciplinary territories.

\subsection{Country distances, niches and aspiration network}

After analyzing the properties of the individual profiles, we now start to compare countries among themselves.
We first display the pairwise distances between national research profiles as a function of the joint productivity of each country pair. This framework is inspired by gravity models, in which the intensity of interaction between two actors is expected to scale with their size and decay with their distance.

\begin{figure}[!h]
\centering
\includegraphics[width=0.95\linewidth]{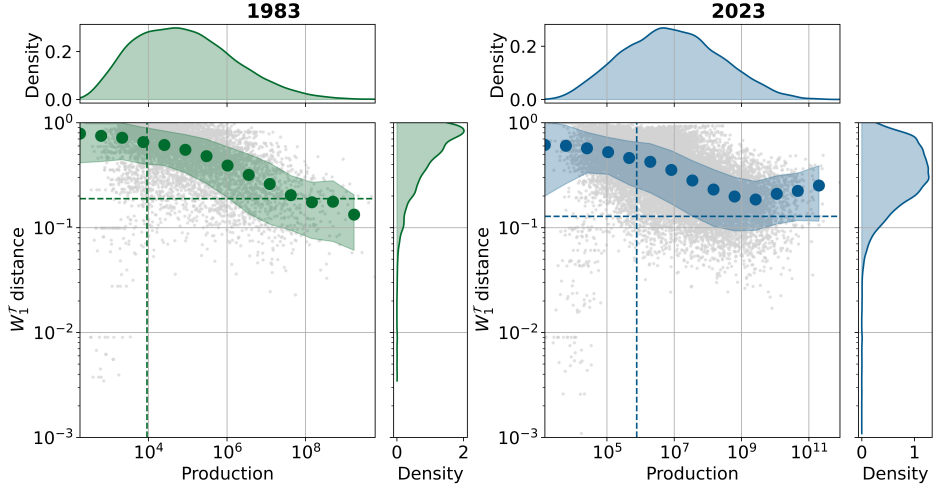}
\caption{Pairwise $W_1^{\mathcal{T}}$ distances between national research profiles as a function of the product of scientific outputs of each country pair, in 1083 and 2023. Each point represents a country pair. Marginal distributions of productivity (top) and distance (right) are shown for each year; colored dotted lines show the lower quartiles of the distributions (used to define the ``niches'').}\label{fig_distributions}
\end{figure}

Fig. \ref{fig_distributions} plots, for each pair of countries, the $W_1^{\mathcal{T}}$ distance between their research profiles against the product of their scientific outputs. In 1983 and 2003, a clear negative relationship is visible: the most productive country pairs (those combining the largest scientific outputs) tend to have the most similar research profiles, clustering at low distances. This pattern is consistent with a world scientific system organized around a dominant core of major producers that share a common disciplinary orientation.
However, by 2023, this relationship had changed markedly. The largest producers are now more distant from each other than in previous decades, suggesting an increasing polarization of the global scientific ecosystem: the major scientific powers are no longer converging toward a shared profile, but are instead developing increasingly distinct research identities.

\begin{figure}[!h]
\centering
\includegraphics[width=0.95\linewidth]{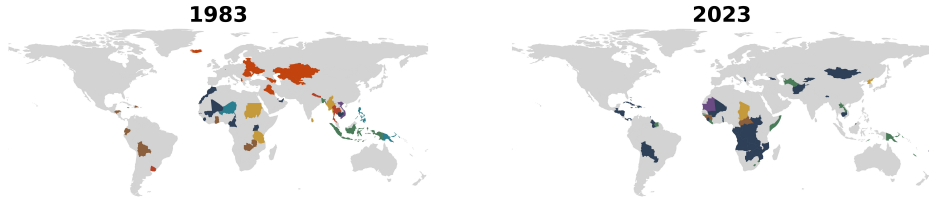}
\caption{Geographic distribution of scientific niches — clusters of countries with small output but highly similar research profiles, in 1983 and 2023. Colors identify distinct niches; grey countries do not belong to any niche.}\label{fig_niches}
\end{figure}

In both periods, the scatter plot reveals the presence of tight clusters of small producers located at very low mutual distances:  countries that, despite limited output, share highly similar research profiles. These scientific niches are shown geographically in the panels of Fig. \ref{fig_niches}. In 1983, the most prominent niche corresponds to the states of the Soviet bloc, whose research profiles were shaped by a common institutional and ideological framework that imposed strong disciplinary homogeneity across the USSR and its allies. By 2023, the geography of scientific niches has shifted dramatically: the tightest clusters now correspond to countries of the Global South, particularly in Latin America and sub-Saharan Africa, whose research profiles converge around a shared set of locally relevant priorities, reflecting both common developmental challenges and the legacy of similar patterns of integration into the global science system.

\begin{figure}[!h]
\centering
\includegraphics[width=0.95\linewidth]{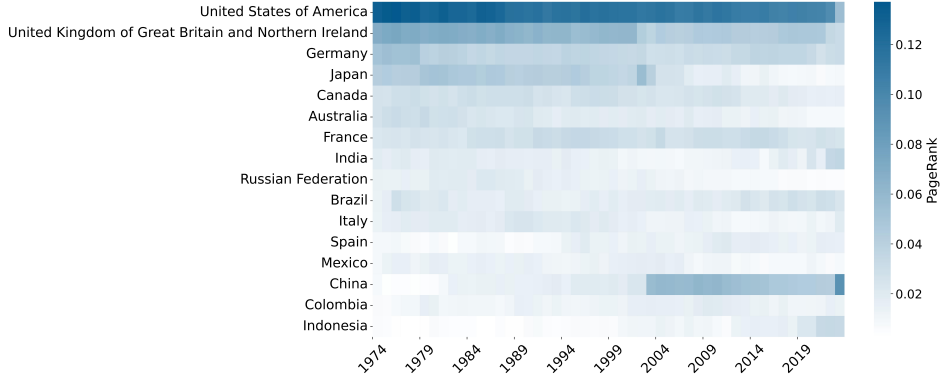}
\caption{Heatmap of PageRank trajectories from 1974 to 2023. }\label{fig_heatmap_rank}
\end{figure}

Fig. \ref{fig_heatmap_rank} introduces the aspiration network, a directed graph in which each country is connected to the $k$ countries (5 in the figure) that simultaneously have a larger scientific output and the most similar research profile: namely, the countries that represent the most plausible and proximate models of scientific development to aspire to. On this directed network, we compute the PageRank centrality, which identifies the countries that are most consistently looked up to as reference models across the global scientific system.

The results reveal a striking reconfiguration of scientific aspirations over the past five decades. The most prominent trend is the rapid rise of China, whose PageRank grows dramatically from the early 2000s onward, reflecting its emergence as a dominant reference model for a large and growing number of countries. Brazil, India, and Indonesia also show increasing PageRank trajectories, signaling their growing role as regional scientific anchors: countries whose profiles are increasingly emulated by smaller producers in their geographic and developmental neighborhood. Conversely, Japan, Russia, and several major European countries display a declining PageRank, suggesting that their role as reference models for the rest of the world has eroded over time, even as their absolute scientific output has remained substantial.

\begin{figure}[!h]
\centering
\includegraphics[width=0.95\linewidth]{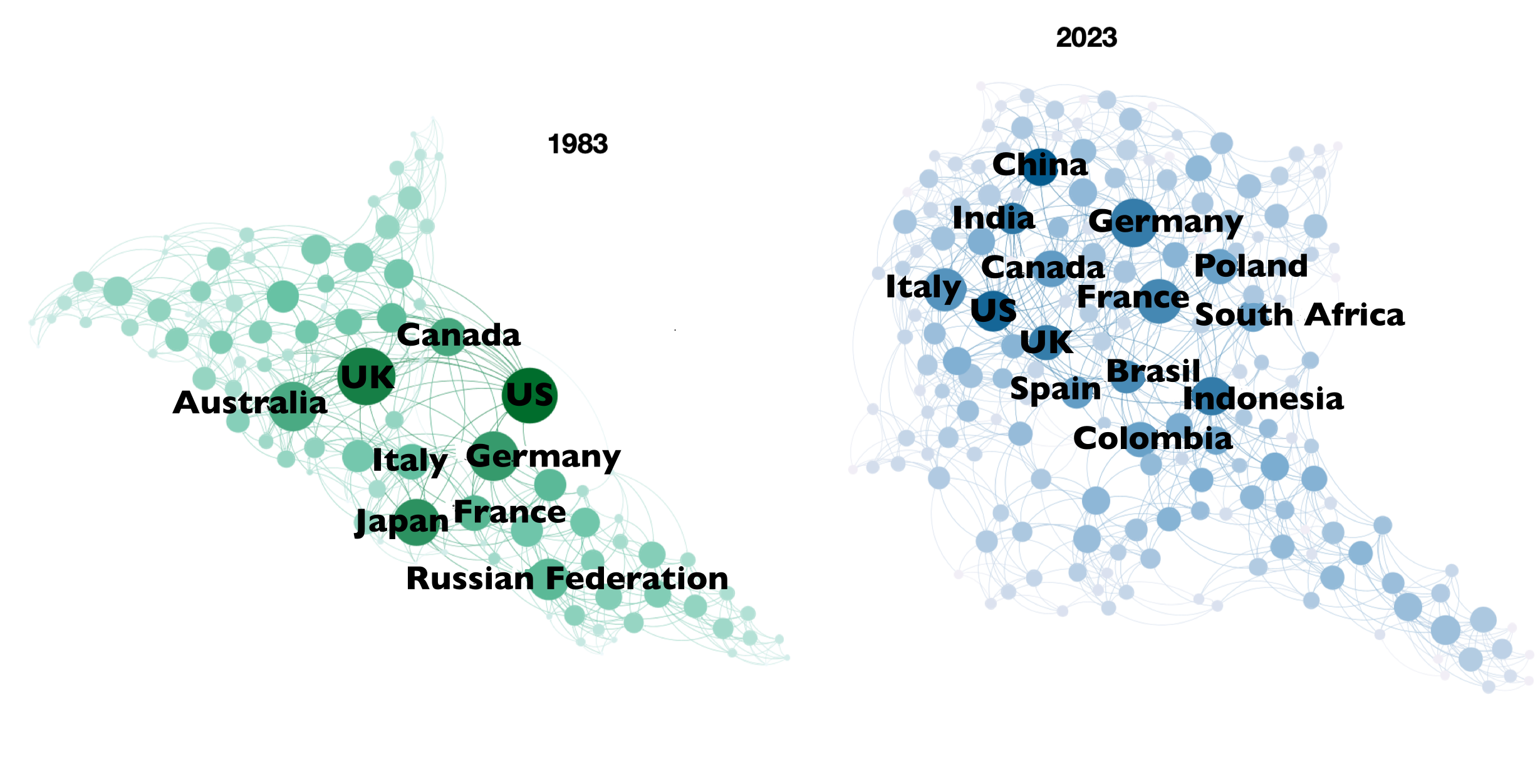}
\caption{Snapshots of the aspiration network in 1983 and 2023. Node size is proportional to in-degree and color intensity is proportional to PageRank centrality.}\label{fig_aspiration_network}
\end{figure}

The two network snapshots for 1983 and 2023 (Fig. \ref{fig_aspiration_network}) corroborate this picture visually. In 1983, PageRank is highly concentrated among a small number of Western powers, with the network organized around a tight and clearly dominant core. By 2023, the distribution of PageRank has become markedly more diffuse, with a larger number of countries achieving significant centrality: a sign that the global scientific system has moved from a unipolar structure, organized around a handful of Western reference models, toward a more multipolar landscape in which multiple regional scientific powers compete for aspirational influence.

\subsection{Mesoscale structures}

\begin{figure}[h]
\centering
\includegraphics[width=0.95\textwidth]{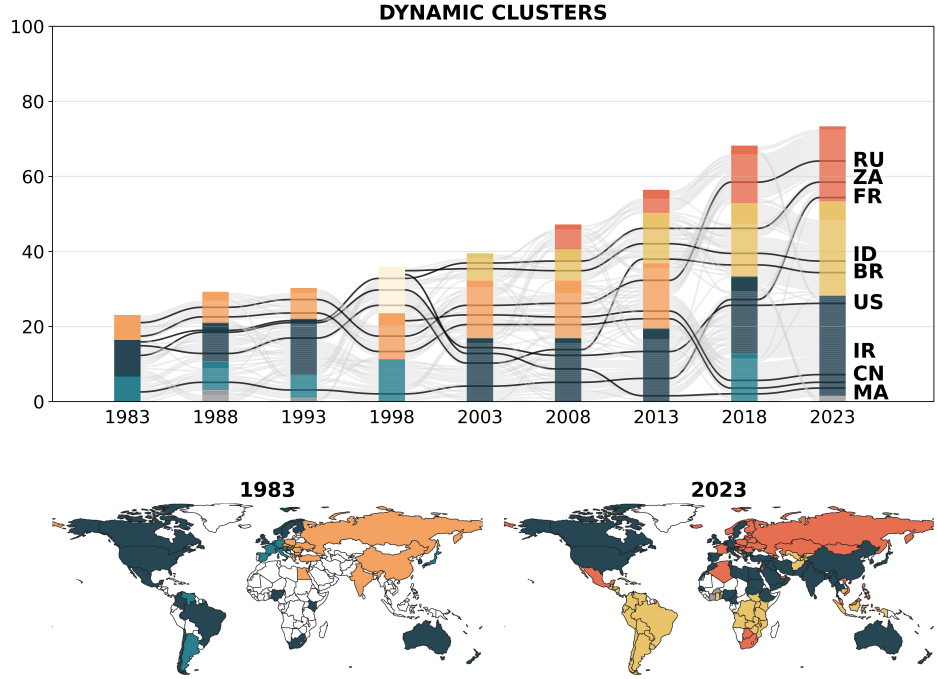}
\caption{Cluster structure of the similarity network every 5 years. Black lines trace the trajectories of selected countries across clusters over time. Maps show the geographic cluster composition in 1983 and 2023; colors are consistent between the maps and the corresponding bars in the chart.}\label{fig_clusters}
\end{figure}

In this section we analyze the temporal evolution of similarity clusters, namely, for each time interval, the groups of countries that share a similar scientific agenda. 
The evolution of country-level clustering in global scientific output, reported in Fig.\ref{fig_clusters}, Fig.\ref{fig_specialization}, reveals a profound transformation in the geopolitical organization of knowledge production over the past four decades. Three distinct phases can be identified.

\subsubsection*{Cold War geography (1983–mid-1990s)}
\begin{figure}[h]
\centering
\includegraphics[width=0.95\textwidth]{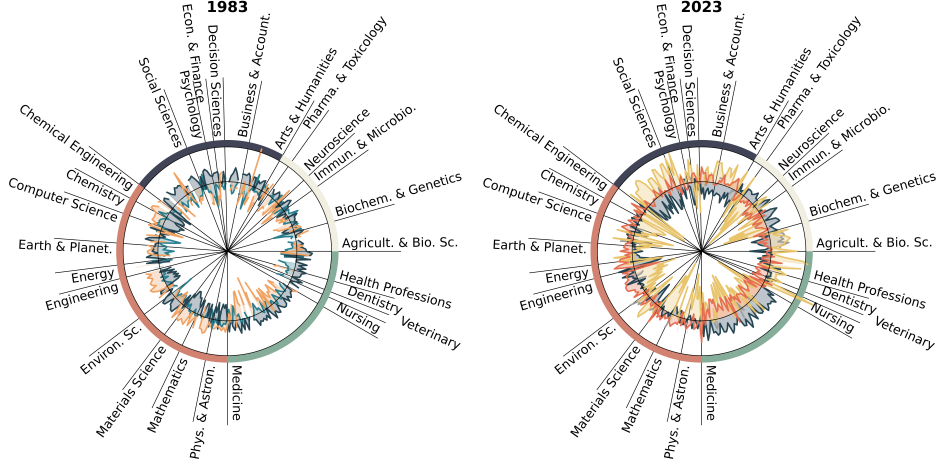}
\caption{Disciplinary specialization profiles of each cluster in 1983 and 2023. Each axis represents a scientific field, and the colored areas show the relative specialization of each cluster. Colors correspond to those used in the maps and bar chart in Fig. 9.}\label{fig_specialization}
\end{figure}

In the early period, the global structure of scientific collaboration closely reflects the bipolar geopolitical order of the Cold War. Three clearly delineated clusters are observable. The first is anchored by the United States and exhibits a clear specialization in medicine, life sciences, and social sciences. The second, linked to the Soviet Union, encompasses China and India alongside Soviet-bloc countries and is characterized by a marked concentration in physics, mathematics, and engineering. A third, distinct cluster groups European countries, which --- unlike the two preceding clusters --- display no strong overall specialization, but show relatively elevated output in selected subfields of biochemistry and the social sciences. These patterns are confirmed quantitatively by the specialization index defined in Eq.\ref{spec_index}, whose values by discipline for 1983 are reported in Fig.\ref{fig_specialization}.

\subsubsection*{Post-Cold War realignment and emerging polarization (late 1990s–2007)}
 
From the early 2000s, the cluster configuration undergoes a significant reorganization. European countries progressively align with the United States, consolidating a broad Western scientific community. A new cluster begins to emerge around Indonesia, specialized in social sciences and agricultural sciences, signaling the growing presence of Southeast Asian countries in global scientific production. China and India remain within the Russia-linked cluster. Crucially, this period witnesses an intensification of epistemic polarization between the two main blocs: the US-centered cluster deepens its specialization in medicine and biochemistry, while the Russia-centered cluster becomes increasingly concentrated in physics and engineering.
 
\subsubsection*{Phase 3: The rise of a Middle Eastern cluster and its shifting alignments (2008–2018)}
  
Beginning around 2008, a cluster of Middle Eastern countries emerges as a distinct actor in the global scientific landscape. By 2013, this cluster aligns with Russia, China, and India, forming an enlarged non-Western scientific bloc. However, this configuration proves unstable: the constituent countries follow diverging disciplinary paths, with China, India, and Middle Eastern countries expanding from physics and engineering into medicine and life sciences, while Russia and post-Soviet countries deepen their engagement with the social sciences. These shifts erode the cohesion of the bloc, and by 2018, China, India, and Middle Eastern countries detach from Russia, and by 2023 their convergence with medicine and life sciences draws them into the orbit of the US-centered cluster.

\subsubsection*{Phase 4: A reconfigured global landscape (2023)}
    
By 2023, the thematic structure of the global scientific ecosystem has changed fundamentally. The axes of differentiation are no longer primarily those of the Cold War. Two main dimensions of separation are now observable. The first is a North–South axis, along which a coherent and growing cluster has consolidated around Brazil and Indonesia. This grouping exhibits the sharpest disciplinary signature of any cluster, with strong over-representation in social sciences, agricultural sciences, veterinary medicine, and environmental studies. The remaining two clusters retain a recognizable thematic division, although both have evolved. The Russia–Europe cluster, while retaining its mathematics and engineering core, indicates a secondary profile in business, arts and humanities, and the social sciences, producing a comparatively broad and less sharply defined specialization. The US-centered cluster --- now encompassing China, India, and Middle Eastern countries --- combines a dominant orientation toward medicine and biochemistry with an increasingly strong engineering component, as the integration of Asian scientific systems has reinforced precisely the disciplinary strengths they carried upon entry. The disciplinary landscape of 2023 thus bears only a faint imprint of the Cold War order. What remains is not a political legacy but a genuine reorganization of the world's scientific priorities.

\section{Data}

Our data are extracted from the OpenAlex 2023 snapshot. In the OpenAlex database papers are associated with at most three topics; we classify each paper according to its primary topic. Topics follow a hierarchical taxonomy: a tree structure comprising 4,516 topics, 252 subfields, 26 fields, and 4 domains (Life Sciences, Health Sciences, Physical Sciences, and Social Sciences). All analyses are conducted at the subfield level (hereafter referred to as "disciplines"), in order to avoid statistical noise associated with the sparse representation of some topics.

Each paper is then assigned to the countries of affiliation of its authors, without duplicates: each country is counted at most once per paper regardless of the number of authors affiliated with it, as we are interested in the presence of a country in a paper rather than its relative contribution. We subsequently filter the dataset to retain only papers for which the affiliation is reported for all authors, ensuring that the geographic attribution is complete and unambiguous.

Since institutional affiliations are resolved to geographic coordinates and then mapped onto present-day national borders, papers affiliated with institutions located in countries that no longer exist (e.g., the Soviet Union, Czechoslovakia, or Yugoslavia) are attributed to the country that currently occupies that location, rather than to the historical political entity. This choice keeps the country definition consistent across the entire time span of the dataset, but it also means that the historical existence of these entities is not preserved in our analysis: an institution that was affiliated with the USSR in 1985, for instance, is counted under Russia, Ukraine, or another successor state depending on its present-day location.

We further filter the dataset in order to only contain papers published between 1970 and 2023.
After these filters our corpus contains around 80 millions of publications, associated to one or more countries and to a scientific discipline. 

\section{Methods}
\subsection{Scientific profiles}
We first associate to each country $i$ its annual number of \textit{domestic} publications, $D_i(t)$, --- papers where all authors are affiliated with the same country --- and its total number of publications, $N_i(t)$, --- papers where at least one author is affiliated with the country.
The fraction of domestic production is therefore defined as $f_i(t)=D_i(t)/N_i(t)$.

For each of the $M$ disciplines, we calculate the annual number of domestic papers produced in discipline $\alpha$ by country $i$, denoted $d_{i\alpha}(t)$.
We define the scientific profile of a country as the vector of shares of its domestic production across disciplines:
\begin{equation}
    \sigma_{i}(t) = \left(\frac{d_{i1}(t)}{D_i(t)}, \ldots, \frac{d_{iM}(t)}{D_i(t)}\right)
\end{equation}

We define the aggregate ``world scientific profile'' as the vector of shares of global domestic production across disciplines:
\begin{equation}\label{worldprofiles}
    \bar{\sigma}(t) =\left(\frac{\sum_i d_{i1}(t)}{\sum_iD_i(t)},\ldots, \frac{\sum_i d_{iM}(t)}{\sum_iD_i(t)}\right)
\end{equation}

For the temporal correlation of world profiles we compare directly the yearly data, since our focus is to identify precise tipping points. 
At the country level, we aggregate the disciplinary production on a rolling time window of five years, in order to have a higher statistical significance for less productive countries. 
In general, in the box plots with the metrics' distributions, we only include in the boxplot calculation the countries with $D_i(t)>500$.

\subsection{Jensen-Shannon and tree-Wasserstein distance}
To measure the dissimilarity between scientific profiles, we use two different distances: the Jensen-Shannon (JS) divergence and the tree-Wasserstein distance ($W_{1}^{\mathcal{T}}$). 
Both the distances treat the profiles as plain probability distributions.
The JS divergence does not have any notion of proximity between disciplines. The $W_{1}^{\mathcal{T}}$, by contrast, incorporates the geometry of the discipline space: two countries are considered closer if their production is concentrated in similar disciplines, even if not identical ones. While the JS divergence is more standard and easier to compute, the $W_{1}^{\mathcal{T}}$ distance is more informative when disciplines have a meaningful structure among them.

The JS divergence is defined as:
\begin{equation}
    \text{JSD}(\sigma_i \| \sigma_j) = \frac{1}{2}D_{\text{KL}}\!\left(\sigma_i \,\Big\|\, \frac{\sigma_i + \sigma_j}{2}\right) + \frac{1}{2}D_{\text{KL}}\!\left(\sigma_j \,\Big\|\, \frac{\sigma_i + \sigma_j}{2}\right)
\end{equation}
and $D_{\text{KL}}(p\|q) = \sum_\alpha p_\alpha \log(p_\alpha / q_\alpha)$ is the Kullback-Leibler divergence. 
To avoid numerical issues arising from zero entries in the profiles, which would make the KL divergence undefined, we apply Laplace (additive) smoothing with $\epsilon = 10^{-9}$ to each component, renormalizing the vector to sum to one afterward.

The tree-Wasserstein distance \cite{evans2011phylogenetic} is defined on a fixed phylogenetic tree $\mathcal{T}~=~(V, E)$ of scientific disciplines. The leaves of the tree $\mathcal{T}$ correspond to individual disciplines, and each country's production profile $\sigma$ is supported on these leaves.

For any node $v \in V$, let $w_v$ denote a length of the edge connecting $v$ to its parent, for the root-node we suppose $w_{\text{root}} = 0$. Then the tree-Wasserstein distance between two profiles is defined as 
\begin{equation}\label{w1-tree-metric}
    W_{1}^{\mathcal{T}}(\sigma_i, \sigma_j) = \sum_{v \in V} w_v |\sigma_i(\Gamma(v)) - \sigma_j(\Gamma(v))|,
\end{equation}
where $\Gamma(v)$ is a set of nodes in the subtree rooted at $v$, and $\sigma(\Gamma(v)) = \sum_{l\in \Gamma(v)} \sigma(l)$ represents the total mass allocated on the leaves of $\Gamma(v)$.

To ensure comparability between settings, we choose the edge lengths $w_v$ so that the maximum possible distance equals 1. All nodes at the same depth share the same edge length. The closer the node is to the root, the bigger its distance to the parent node is. This design encodes the structure of scientific knowledge: two disciplines within the same field are penalized less than two disciplines from different fields, which are in turn penalized less than disciplines from entirely different domains.

Empirical simulations show that pairwise distances between profiles stabilize once each sample reaches approximately 500 articles. Below this threshold distances are overestimated. We adopt this threshold as a cutoff for inclusion in the analysis, accepting a small upward bias for close profiles in exchange for broader country coverage.

\subsection{Epistemic Entropy measures}
We define the epistemic entropy of a country $i$ as the Shannon entropy of its profile $\sigma_i(t)$:
\begin{equation}
    H_i(t) = -\sum_{\alpha=1}^{M} \sigma_{i\alpha}(t) \log \sigma_{i\alpha}(t)
\end{equation}
which measures the diversification of the domestic production across disciplines.

While $H_i(t)$ captures the degree of diversification of the national scientific profile, it treats all subfields as equidistant, disregarding the taxonomic structure of the discipline space. To account for the relational geometry of scientific knowledge, we introduce a complementary measure: the \textbf{Wasserstein Epistemic Entropy} of country $i$ at time $t$ as:

\begin{equation}
    \mathcal{H}_i(t) = W_1^{\mathcal{T}}\!\left(\sigma_i(t),\, u\right)
\end{equation}

where $u = (1/M, \dots, 1/M)$ is the uniform distribution over subfields. $\mathcal{H}_i(t)$ vanishes if and only if $\sigma_i(t) = u$, and is large when production is concentrated on disciplinarily proximate subfields, i.e.\ subfields sharing a common ancestor close to the leaves of $\mathcal{T}$.

$H_i(t)$ and $\mathcal{H}_i(t)$ are not redundant: two profiles with identical Shannon entropy but different disciplinary distributions will in general yield different values of $\mathcal{H}_i(t)$, with higher values assigned to profiles concentrated within a single branch of the tree.

\subsection{Aspiration network}
Based on the $W_{1}^{\mathcal{T}}$ distance, we compute a ``scientific distance'' matrix among countries, $\Delta_{ij}$.

For each country $i$, we define its ``aspiration'' toward country $j$ as:
\begin{equation}
    A_{i\rightarrow j}(t) = \frac{\log D_j(t)}{\Delta_{ij}^\beta}\,\theta\!\left(D_j(t) - D_i(t)\right)
\end{equation}
where $\theta(\cdot)$ is the Heaviside step function, ensuring that country $i$ only aspires toward countries with a larger domestic production, and $\beta$ is a parameter controlling the decay of aspiration with scientific distance. 
We fix $\beta = 0.5$ in order to balance the influence of scientific distance and country size, avoiding an excessive penalization of scientifically distant countries.

We then define the \textit{aspiration network} as a directed graph in which each country $i$ is connected to its top-$k$ most aspiring countries, i.e.\ the $k$ countries $j$ with the highest aspiration score $A_{i\rightarrow j}(t)$. In the following, we set $k=5$.

\subsection{Clustering methods}

\subsubsection{Static Clusters}
For every year, from the pairwise distance matrix $\Delta_{ij}$, we derive a similarity matrix using a Gaussian kernel: 
\begin{equation}
\Sigma_{ij} = \exp \left\{-\dfrac{\Delta_{ij}^2}{2 \gamma^2}\right\},    
\end{equation}

where $\gamma=0.1$. With this choice, pairs of countries whose scientific distance exceeds 0.5 contribute a similarity below $e^{-1/2} \approx 0.07$, effectively suppressing distant connections and producing a sparser network structure.

We apply the modularity-based Leiden algorithm \cite{traag2019louvain} to the similarity network to detect communities --- groups of countries sharing similar scientific profiles. A key advantage of this approach is that it requires neither a predefined number of clusters nor a target inter-community distance, avoiding a potential source of arbitrary choices in the analysis.

The clustering results exhibit relatively low modularity (0.2--0.3) and silhouette scores (0.3--0.4), suggesting weak community boundaries. Furthermore, since the Leiden algorithm is sensitive to the initial assignment of nodes, countries near cluster borders may be inconsistently classified across runs. To address both issues, we use evidence accumulation \cite{ana2005evidence}: run the algorithm 30 times with independent initializations and assess the consistency of the resulting partitions using the Adjusted Rand Index (ARI).

The median ARI in the 50-year analysis window is 0.87, confirming that the detected community structure is largely consistent across initializations. We then construct a co-occurrence matrix, where each entry records the fraction of runs in which two countries are assigned to the same cluster. The distribution of co-occurrence values is strongly bimodal, with most pairs of countries either never or almost always clustered together, reflecting well-separated cores with ambiguity concentrated near cluster boundaries. We use this co-occurrence matrix as the input for a final run of the Leiden algorithm, producing the definitive community assignments.

\subsubsection{Dynamic Clusters}
The full procedure of community detection is repeated independently for each year. To track how communities evolve over time \cite{greene2010evolution}, we represent each detected cluster by its aggregated scientific profile 

\begin{equation}\label{cluster_profile} 
\hat\sigma_C(t) = \left( \dfrac{\sum_{i \in C}\sigma_{i1}(t)}{\sum_{i \in C, \alpha \in \overline{1, M}}\sigma_{i \alpha}(t)}, \dots , \dfrac{\sum_{i \in C}\sigma_{i M}(t)}{\sum_{i \in C, \alpha \in \overline{1, M}}\sigma_{i\alpha}(t)}\right)
\end{equation}

and compare these profiles across years using the following procedure.

Clusters identified in the first year of analysis are registered as the initial set of dynamic clusters, each uniquely determined with an aggregated profile.

For each subsequent year, every detected cluster is compared against all existing dynamic cluster profiles. If the tree-Wasserstein distance between a new cluster and an existing dynamic cluster falls below 0.1, the new cluster is identified as a continuation of that dynamic cluster, and the corresponding profile is updated with the new aggregate. Dynamic clusters that find no continuation in a given year are not updated, but their profiles are retained for comparison in future years, allowing for the possibility of reemergence.

The boundary is equal to the median distance observed across countries inside clusters.

This procedure models the inheritance and evolution of scientific communities over time. By anchoring cluster identity to aggregated disciplinary profiles, gradual shifts in research priorities are naturally captured: changes that are subtle within a single year may accumulate into notable reorientations over longer horizons.

\subsubsection{Dynamic Cluster Specialization}

Dynamic clustering is performed on the basis of the proximity of countries' scientific profiles. To interpret the resulting structure and its evolution over time, we analyze the aggregated scientific profiles of each cluster.

For each year and each cluster, we measure the representation of scientific disciplines among the cluster's member countries relative to the global average, adjusting for sample size using finite population correction. 
Let $\log \sigma_{i\alpha}(t)$ denote the log-transformed profile value of country $i$ in discipline $\alpha$ at year $t$, we observe that this quantity is normally distributed across countries.
We denote by $\{\mu_\alpha, s_\alpha, n_\alpha\}$ the mean, standard deviation and sample size of $\log \sigma_{i\alpha}(t)$ across all countries in a given year, and by $\{\mu_{C\alpha}, n_{C\alpha}\}$ the mean and sample size  restricted to the countries belonging to cluster $C$. The specialization index of cluster $C$ in discipline $\alpha$ \cite{oliveira2025overepresentation} is then defined as:

\begin{equation}\label{spec_index}
    SI_{C \alpha} = \dfrac{\mu_{C \alpha} - \mu_\alpha}{\sqrt{\dfrac{n_\alpha - n_{C \alpha}}{n_\alpha - 1} \times \dfrac{s_\alpha^2}{n_{C \alpha}}}}.
\end{equation}

This index measures how far the cluster's mean log-profile deviates from the global mean, expressed in units of the finite-population-corrected standard error. Positive values indicate that discipline $\alpha$ is over-represented within the cluster relative to the global average, while negative values reflect under-representation. Values close to zero indicate a representation consistent with the global average.

The index is sensitive to cluster size: larger clusters yield a smaller standard error, so that even modest deviations from the global mean produce larger index values, lending greater statistical weight to well-populated clusters.

The specialization index serves as a key interpretive tool throughout the analysis: it allows us to identify the disciplines that characterize each cluster, to better understand the emergence and dissolution of clusters in the dynamic setting, and to explain why individual countries shift from one cluster to another over time.

\section{Discussion}
In this article, we have analyzed the geopolitical structure of scientific knowledge from several perspectives.
We first positioned ourselves at the global level, examining the sensitivity of worldwide research directions to major geopolitical events, in order to situate ourselves within the debate between cumulative and disruptive scientific evolution. We highlight two main regimes: a slow, cumulative evolution, with a certain tendency to accelerate over the past 20 years, and significant shocks that have disrupted this trajectory. Notably, all of these tipping points are exogenous in nature: they were not driven by particularly innovative papers or research subjects, but by external factors: Chernobyl, 9/11, and Covid.

We then examine whether the scientific system follows the globalization dynamics observed in the economic and cultural world from the 2000s onward. We find, first of all, that a trend toward internationalization of collaborations is clearly present at a large scale. However, going against this trend, countries experiencing scientific growth tend primarily to consolidate a national research system. Looking more closely at scientific agendas, we observe that while the major producing countries tend to align themselves increasingly with the global agenda, the national system-building dynamics of emerging countries are in some cases accompanied by a genuine specialization in specific disciplines that fall outside the global agenda.

Finally, we examine the sources of scientific inspiration across countries. We find that contemporary science has a less polarized structure than in the 1980s, with more diversified sources of inspiration.
The structure in terms of scientific clusters likewise reveals a clear reconfiguration, with a significant emerging cluster that appears to aggregate, around rising nations such as Brazil and Indonesia, the scientific agendas of the Global South.

This study aimed to examine the geopolitical structures present in the scientific ecosystem, drawing not so much on recognition in terms of citations, but adopting the perspective of scientific production itself. The dynamics of scientific impact deserve a more thorough treatment, which will be developed in future work.

Another central dynamic in this regard is the mobility of the scientific workforce. Already analyzed in previous work, it would merit being revisited and deepened in light of the observations concerning scientific agendas.

A further limitation is the absence of scientific funding systems from our analysis. In OpenAlex, the number of papers associated with a funder is very limited. We are, however, working to acquire this type of information by cross-referencing different datasets (for example Web of Science), and this line of analysis will underpin future work.

Finally, a growing literature from both STS and decolonial theory points to clear inequalities in research practices within international collaborations. More specifically, the hierarchical division of scientific labor often found in North–South collaborations, in which Southern partners are asked to primarily contribute empirical material, while research questions and theoretical frameworks are formulated in the North, may help explain, at least in part, the phenomenon of extraversion observed in our data. As scientific production in the Global South becomes increasingly internationalized, it simultaneously tends to diverge from domestic research priorities, suggesting a growing heteronomization of local intellectual resources and their mobilization in support of research agendas defined in the North. This is certainly a topic of great interest and would merit, in the future, in-depth qualitative fieldwork comparing domestic and international research practices in some of the countries that this quantitative study identifies as particularly interesting, for example, the Central African countries that align with the Brazil-Indonesia cluster.

\backmatter

\begin{appendices}






\end{appendices}


\bibliography{sn-bibliography}

\end{document}